# Strain-Engineered High Responsivity MoTe₂ Photodetector for Silicon Photonic Integrated Circuits

R. Maiti[1], C. Patil[1], T. Xie[1], J.G. Azadani[2], M.A.S.R. Saadi[3], R. Amin[1], M. Miscuglio[1], D. Van Thourhout[4], S.D. Solares[3], T. Low[2], R. Agarwal [5], S. Bank [6], V. J. Sorger[1]*

[1]Department of Electrical and Computer Engineering, George Washington University, Washington, DC 20052, USA

[2]Department of Electrical and Computer Engineering, University of Minnesota, Minneapolis, MN 55455, USA

[3]Department of Mechanical and Aerospace Engineering, George Washington University, Washington, DC 20052, USA

[4]Department of Information Technology, Ghent University - IMEC, Technologiepark Zwijnaarde 126, 9052 Gent, Belgium

[5]Department of Materials Science and Engineering, University of Pennsylvania, Philadelphia, PA 19104, USA

[6]Department of Electrical and Computer Engineering, University of Texas, Austin, TX 78758, USA

*Corresponding Author E-mail: sorger@gwu.edu

## Abstract

In integrated photonics, specific wavelengths are preferred such as 1550 nm due to low-loss transmission and the availability of optical gain in this spectral region. For chip-based photodetectors, layered two-dimensional (2D) materials bear scientific and technologically-relevant properties such as electrostatic tunability and strong light-matter interactions. However, no efficient photodetector in the telecommunication C-band has been realized with 2D transition metal dichalcogenide (TMDCs) materials due to their large optical bandgap. Here, we demonstrate a MoTe₂-based photodetector featuring strong photoresponse (responsivity = 0.5 A/W) operating at 1550 nm on silicon micro ring resonator enabled by strain engineering of the transition-metal-dichalcogenide film. We show that an induced tensile strain of ~4% reduces the bandgap of MoTe₂, resulting in large photo-response in the telecommunication wavelength, in otherwise photo-inactive medium when unstrained. Unlike Graphene-based photodetectors relying on a gapless band structure, this semiconductor-2D material detector shows a ~100X improved dark current enabling an efficient noise-equivalent power of just 90 pW/Hz$^{0.5}$. Such strain-engineered integrated photodetector provides new opportunities for integrated optoelectronic systems.

**Keywords:** Integrated Photonics, Tensile strain, KPFM, work function, TMDCs, photodetector, microring resonator

## Introduction

Strain engineering of traditional semiconductors like Si/Ge and III-V semiconductors can be utilized to enhance the performance of electronic and photonic devices [1-3]. By inducing strain, the electronic band structure can be modified by epitaxial growth techniques to control the lattice constant, which, for example, can enable a reduction of the effective mass and, thus, positively impacting mobility [4,5]. Lowering the dimensionality from bulk crystals to 2D layered films, can enable the material to sustain higher amounts of strain. A straightforward method to achieve strain in 2D nanocrystals is through mechanical bending, such as transferring them onto flexible substrates or wrapping them around a pre-patterned structure [6-9]. Under small (<2%) compressive (tensile) strain, the bandgap increases (decreases) and can even induce a semiconductor-to-metal phase transition (10% for monolayer) in $MoS_2$ [10]. Furthermore, a strain-induced exciton redshift and exciton funnel effect are present in few-layer $MoS_2$ crystals [11,12]. Recently, it was shown that strain can strongly modulate the bandgap energy (~70 meV) of monolayer $MoTe_2$ and $MoWTe_2$ for 2.3% of uniaxial strain [13]. Beyond such pioneering demonstrations, there is however a lack of experimental evidences of strain-induced bandgap engineered optoelectronic devices for the telecommunication wavelength, which is needed to assess the potential of this class of materials as building blocks for the future integrated photonic platform.

An integral device for photonic circuitry supporting a plethora of applications, such as sensing, data communication, and general signal processing is a monolithically integrated photodetector operating at near-infrared (NIR) [14-17]. The wavelength of 1550 nm is a prominent spectral choice since it a) overlaps with the gain spectrum of erbium-doped-fiber-amplifiers, and b) is transparent for foundry-based silicon photonics. The current state-of-the-art NIR photodetectors utilize InGaAs, InP, and Ge due to their high absorption (>90%) at telecommunication wavelengths [18-20]. However, III-V materials are not compatible with Si CMOS technology due to the complexity of growth, wafer bonding issues, and thermal budget. On the other hand, Ge photodetectors typically show higher noise due to the presence of defects and a dislocation center at the Si-Ge interface during the epitaxial growth process [21]. In contrast, heterogeneous integration of 2D materials with photonic platforms bears several advantages including strong exciton binding energy, unity-strong index tunability and CMOS compatibility due to lattice matching requirements (weakly bonded van der Waals forces) [22-27]. The prominent example of Graphene-based integrated photodetectors, while being functional has a fundamental challenge to achieve low dark current due to the gapless band structure when operated in the photoconductive mode [23]. On the other hand, black phosphorus (BP), a 2D nanocrystal of phosphorus, shows high responsivity and low dark current at 1550 nm [20]. However, the low stability under ambient conditions, limits its application [28].

In this work, we demonstrate a strain-engineered photodetector based on heterogeneously integrating a multi-layer (ML) 2H-MoTe$_2$ crystal flake atop a silicon microring resonator (MRR). Straining the ML MoTe$_2$ red-shifts the bandgap from 1.04 eV down to 0.8 eV, thus enabling absorption at the photonic integrated circuit (PIC)-relevant wavelength of 1550 nm. Intentionally wrapping a MoTe$_2$ nanocrystal around a non-planarized waveguide induces local tensile strain overlapping with the waveguide's optical mode. We measure a photo responsivity of 10 mA/W (40nm thick) and 0.5 A/W (60 nm thick) at -2 V. The device shows a low dark current of just 13 nA, a noise equivalent power (NEP) of 90 pW/Hz$^{0.5}$, and operates up to ~35 MHz. Such strain-engineered photodetectors using 2D-materials that are integrated with established photonic platforms could potentially open up a new class of optoelectronic components.

**Results & Discussions:**

Continuous tuning of physical properties by controlling the mechanical deformation, such as strain, offers possibilities for significantly modifying both the electronic and photonic properties of 2D materials [29, 30]. Here, our approach is to exploit strong local uniaxial strain towards reducing the optical bandgap of ML MoTe$_2$ nanocrystals co-integrated onto Si photonic waveguide-based structures. As an example, we show that such setup enables photodetection at the technologically-relevant wavelength of 1550 nm (Fig. 1). Bulk MoTe$_2$ is an indirect gap semiconductor with a bandgap of 1.04 eV, where the conduction band minimum lies along K-Γ̄ symmetry line and the valence band maximum is located at the K point. Naturally, pristine ML (non-strained) MoTe$_2$ is associated with low absorption at 1550 nm resulting in low photo responsivity of only a few mA/W due to a roll-off in the absorption edge, as demonstrated by a flat MoTe$_2$ atop a planarized waveguide (Supplementary online information, section 6, Fig. S6). However, intentionally wrapping the 2D film around a non-planarized waveguide of height ~220 nm induces a localized tensile strain near the waveguide (Fig. 1a,b). Fundamentally, the performance of 2D material-based photonic devices is determined by the ability of the mode of optical waveguide to interact with the 2D nanocrystal. One typical approach to circumvent this limitation is to place the 2D film onto a planar waveguide and evanescently couple to the optical mode so that optical interaction length is not only dictated by the 2D film thickness, but rather by its longitudinal length [31, 32]. Such brute-force device engineering, however, results in sizable device footprints, thus negatively impacting the electrical device performances, for example the energy-per-bit efficiency and RC response time, which are both adversely affected by increased electrical capacitance. To reduce the footprint and improve electrical performance, here, we integrate 2D nanocrystals with microring resonators (MRR), thus increasing the weak light-matter interaction (Fig. 1).

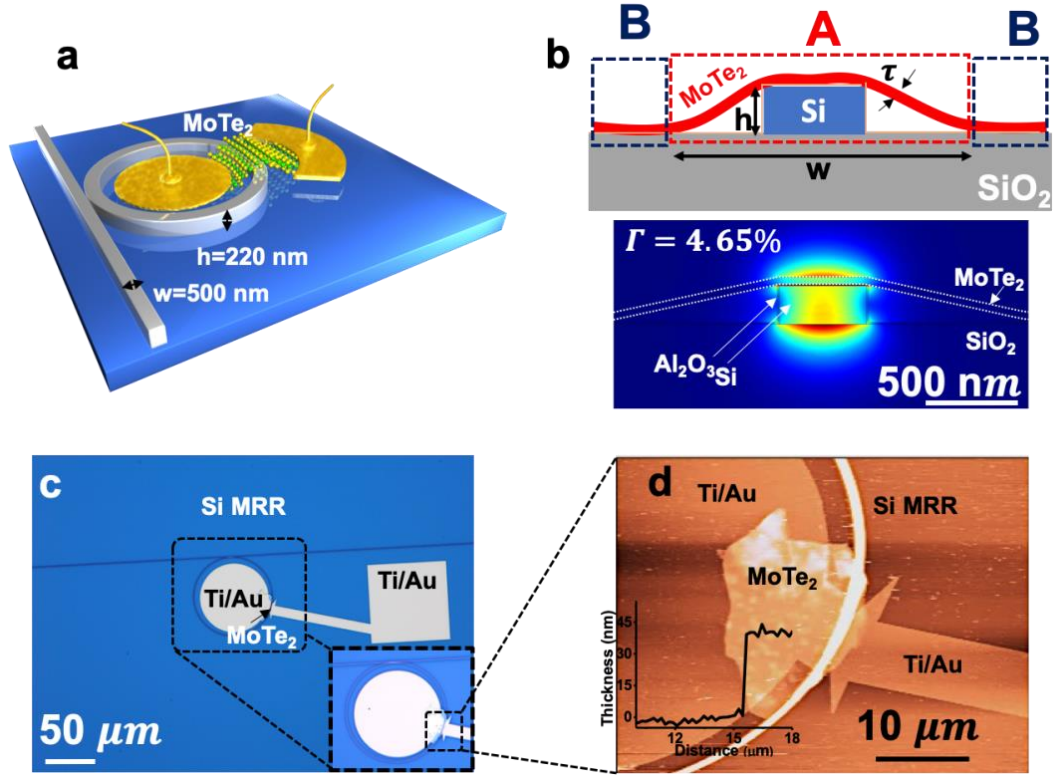

**Figure 1. Microring resonator integrated photodetector** a) Schematic illustration of a microring resonator (MRR) integrated MoTe$_2$ photodetector. b) Schematic diagram of bent 2D nanocrystal on top of non-planarized waveguide (bending width=w, height=h and thickness=τ), introducing strong localized tensile strain in region A (red marked), whereas, the unstrained region is marked as region B (dark blue marked). Simulated mode profile for MoTe$_2$ integrated waveguide for transverse magnetic (TM) mode, where the cross-sectional structure is extracted from an AFM micrograph. c) Optical micrograph of the device (top view) where few layers of MoTe$_2$ nanocrystal are integrated onto a non-planarized Si microring resonator (Radius=40 μm, height h=220 nm and width $w$=500 nm) with a spacing layer of ~10 nm thick Al$_2$O$_3$ by using a 2D printer technique [33]. Ti/Au was deposited as electrical contact pads on both sides of the ring resonator to facilitate efficient collection of photogenerated charge carriers. d) AFM topography of the active device area showing the thickness of the flakes ~40 nm (height profile in inset).

Images of the device illustrate the precise placement capability of exfoliated $MoTe_2$ flakes enabled by utilizing our in-house developed 2D material printer technique (Fig. 1c,d) and Supplementary online information Sections 1 & 2, and Figs. S1 & S2) [33], atop a thin (10 nm) $Al_2O_3$ layer acting as an electrical isolation layer between the silicon on insulator (SOI) photonic chip with the 2D nanocrystal. The optical mode at 1550 nm (Fig. 1b) couples with the $MoTe_2$ layer through the evanescent field, leading to optical absorption and the generation of photo-generated carriers, which will be collected by the two metal electrodes made of Ti (5 nm)/Au (45 nm), contacted on opposite sides of the microring resonator (Fig. 1c). The channel length of this $MoTe_2$ two-terminal photodetector is ~800 nm (waveguide width ~500 nm), where one of the electrical contacts is positioned ~100 nm away from the edge of the ring resonator to create a lateral metal semiconductor metal (M-S-M) junction that overlaps with the waveguide mode (Fig. 1b). Two exemplary devices with different dimensions of the 2D material (coverage length and thickness of the transferred flakes are 15 and 31 $\mu$m and 40 and 60 nm, respectively), atop of ring resonator are discussed next (Fig. 1d).

A representative current-voltage (I-V) curve shows efficient photodetection indicated by the 100:1 photo to dark current ratio at -1 V bias (Fig. 2a). The device is associated with low dark current ~13 nA at -1 V bias, which is about 2-3 orders and ~2 times lower compared to graphene and transition metal dichalcogenides (TMDCs)-graphene contacted photodetectors, respectively [26, 27] (Supplementary online information, section 11, Table S1). The symmetric nature of the I-V curve indicates the formation of two back-to-back (Ti/$MoTe_2$) Schottky junctions. The working principle of this detector is photocarrier generation across the bandgap, where the applied voltage bias across the 2-terminal contacts enables charge carrier separation (Fig. 2b); at equilibrium, the work function (4.3 eV) of titanium ensures Fermi level alignment with the p-doped MoTe$_2$ [26] (Fig. 2 b,i). The formation of the Schottky barrier at the junction suppresses carrier transport, thus resulting in low dark current. With applied bias voltage, the potential drop across the junction reduces the Schottky barrier height (Fig. 2b, ii)). Upon illumination of the laser source, the generated photocarriers separate due to the formation of a built-in potential inside the junction, resulting in a photocurrent. In order to obtain the photo-responsivity, we test the detector's response as a function of waveguide input power and bias voltage (Fig. 2c,d). After calibrating for coupling losses (Supplementary online information, section 10, Fig. S10), we find an external responsivity (i.e. $I_{photo}/P_{input}$) of 10 and 468 mAW$^{-1}$ at -2 V for device 1 (MoTe$_2$ dimensions: thickness = 40 nm, MRR coverage length = 15 μm) and device 2 (thickness = 60 nm, MRR coverage length = 31 μm), respectively, which is 1.75 times higher compared to a waveguide integrated MoTe$_2$ detector tested at 1310 nm [26] (Supplementary online information, section 11, Table S1). The high responsivity of these MoTe$_2$ detectors operating at 1550 nm can be attributed to enhanced absorption from 1) the strain-engineered lowered bandgap, and 2) from the MRR photon lifetime enhancement proportional to the finesse of the cavity (Supplementary online information, section 3, Fig. S3). The responsivity varies linearly as a function of bias voltage, corresponding to a back-to-back (M-S-M) junction and shows that the device is not driven yet into saturable absorption at these power levels (Fig. 2c&d). The external quantum efficiency (EQE) can be determined by, EQE = $R*hc/q\lambda$, where, $R$, $h$, $c$, $q$, and $\lambda$ are the responsivity, Planck constant, speed of light in vacuum, elementary electron charge, and operating wavelength, respectively. The EQE's for these two devices are 1% (device 1) and 37% (device 2) at -2 V (Fig. 2d). The variation of responsivity as a function of optical input power shows a flat response until $P_{in}$=30 μW, where state-filling blockage sets-in as the generation of excess carriers increases the radiative recombination for higher power (P$_{in}$) in the waveguide (Fig. 2e). We first consider the impact of the microring resonator on the detector's performance; when operated at a fixed power of 20 μW at different bias voltages, we find a ~50% enhanced photocurrent ON (vs. OFF) resonance, which matches the MRR's finesse of ~1.6 (Fig. 2f and supplementary online information, section 4, Fig. S4). However, the MRR integrated MoTe₂ photodetector exhibits a ~1.2 times enhancement of the photocurrent at 1550.75 nm (on resonance), as compared to the straight waveguide photodetector (non-planarized) (Fig. 2f). (Supplementary online information, Section 5, Fig. S5). We note that while a MRR with a higher finesse will improve responsivity, yet it can reduce the detector's 3dB response speed due to a longer photon cavity lifetime, if the latter is the limiting factor and not the carrier lifetimes (i.e., relating to the gain-bandwidth product figure of merit of photodetectors).

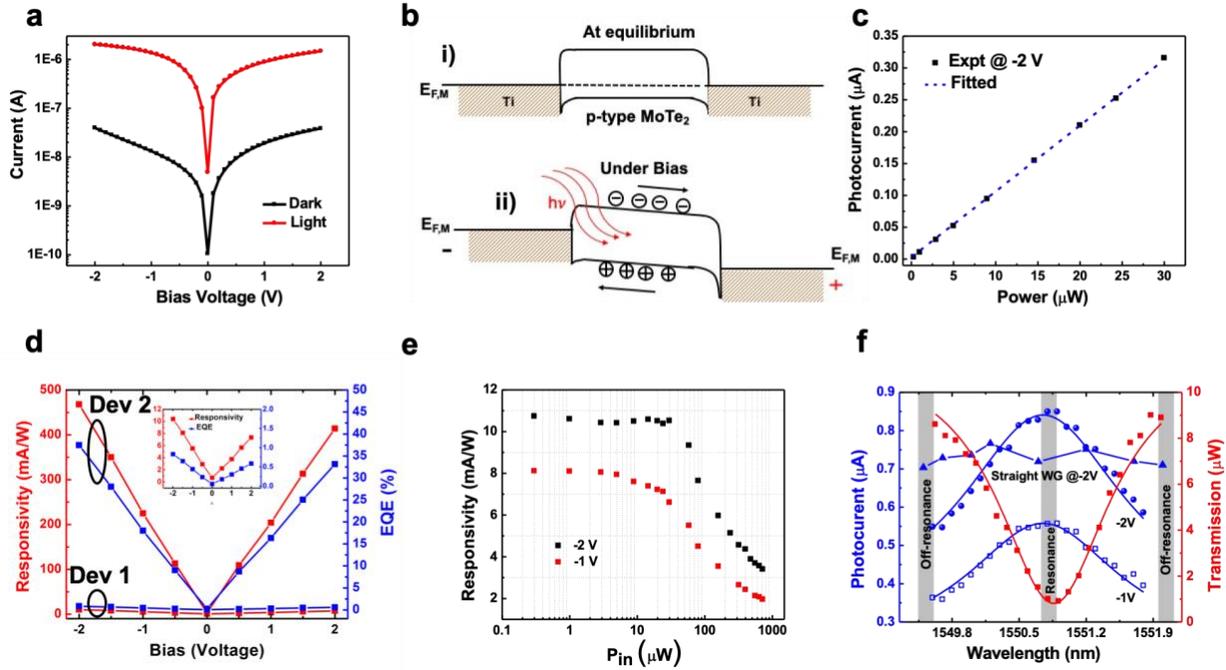

**Figure 2. Photoresponse of the Au/MoTe₂/Au integrated on MRR** a) Typical I-V characteristics (semi-log plot) of Au/MoTe₂/Au diode showing ~2 orders of magnitude enhancement for light conditions (red) as compared to dark (black), b) Schematic energy band diagram explains the photodetection mechanism: i) under equilibrium state, and ii) under bias showing the flow of charge carriers upon excitation. c) Photocurrent vs incident optical power showing the responsivity of the device of ~10.3 mA/W at -2 V, d) Responsivity and external Quantum Efficiency (EQE) as a function of bias voltage for two devices (device 1 thickness-40 nm & coverage length-15 μm and device 2-thickness-60 nm & coverage length-30.7 μm), showing symmetric linear variation due to M-S-M device configurations. Zoomed-in responsivity and EQE plot for device 1 (inset). e) Responsivity of the Au/MoTe₂/Au detector as a function of illuminated optical power for -1 v and -2 v, respectively. f) Spectral response of MRR integrated photodetector showing maximum responsivity at resonance wavelength (1550.75 nm) for -1V (open squares) and -2V (closed spheres) when the optical transmission (closed red squares) is minimum. Photodetector showing ~50% photocurrent enhancement compared to off-resonant conditions (1549.65 nm). Spectral response of MoTe₂ integrated non-planarized straight (not MRR) waveguide is shown in blue closed triangles as a reference.

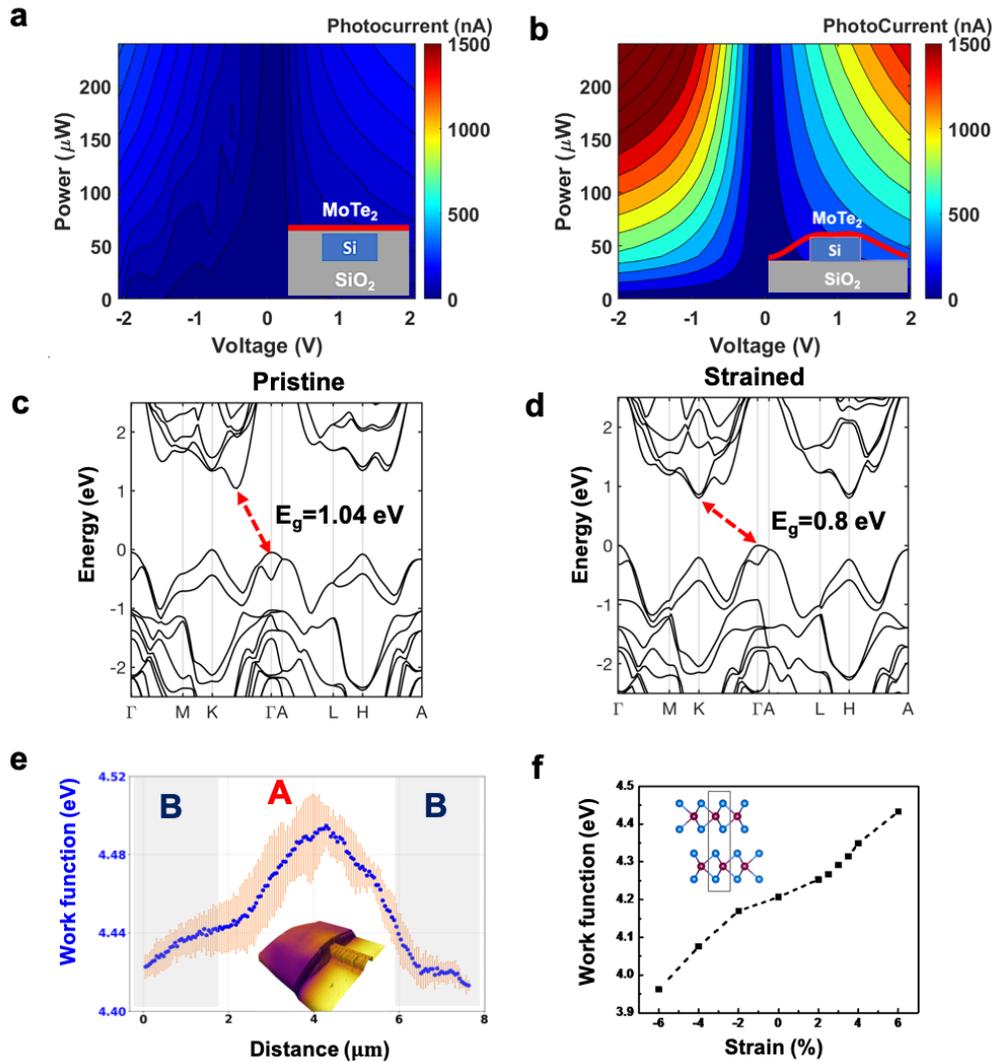

**Figure 3. Mechanism of enhanced photocurrent.** Measured photocurrent as a function of incident light power and electric bias at room temperature for a (a) planarized & (b) non-planarized (schematic shown in inset of Fig. 3a and 3b respectively) MoTe$_2$ photodetector at 1550 nm. Bulk band structure of (c) pristine and (d) 4% strained MoTe$_2$ calculated using DFT showing lowering of the bandgap. the valence band maximum is set to zero. e) An average of several KPFM scan lines across the waveguide. The change in work function is attributed to the strain imparted by the waveguide on the MoTe$_2$ flakes. The non-uniformity in work function in region A is due to the asymmetric nature of the strain in the flakes around the waveguide. A map of work function of the strained flakes obtained by Kelvin probe force microscopy (KPFM) overlaid on topography shown in the inset of figure 3e. It is evident that the work function increases locally in regions where the flakes are strained by the waveguide geometry (on top and in close vicinity of the waveguide- region A in Fig. 1b), compared with the unstrained or flat region (located where the flakes come into contact with the substrate – region B in Fig 1b). f) Variation of work function with different values of strain. Negative and positive strain values correspond to compressive and tensile strain, respectively. Side view of the crystal structure of bulk MoTe$_2$ with its unit cell is shown by inset. Blue and maroon balls represent Te and Mo atoms, respectively.

Strain-induced modulation of the electronic bandgap in 2D semiconductors wrapped around a patterned substrate has previously been observed [34]. Our experimental observation of the enhanced photocurrent for the wrapped around detector versus the planarized control sample might have also benefitted from the induced-strains (Figure 3 a&b, (Supplementary online information, section 6, Fig. S6). Indeed, our calculated band structures by using first-principles density functional theory (DFT) for ML MoTe$_2$ found the electronic gap to acquire a red-shift with tensile strains, potentially bridging the electronic gap with the telecom wavelength (Supplementary online information, section 7, Fig. S7). Subject to higher tensile strain, we note that the valence bands shift towards higher energies (at $\Gamma$). In addition, the conduction bands at the K and H points shift towards lower energies and become the conduction band minimum with equal energies (Fig. 3d). As a result, the bandgap reduces from 1.04 eV for pristine to 0.80 eV for strained MoTe$_2$ (4%), yet the material remains an indirect band gap semiconductor (Fig. 3d). These DFT results suggest that tensile strain can open up interband optical transitions at the telecom frequency, with is otherwise forbidden in the unstrained case. These are consistent with our observed experimental findings of enhanced photoresponse at 1550 nm for strained (vs. pristine) photodetectors discussed above.

To obtain a quantitative picture of strain-induced band structure modulation, we performed Kelvin probe force microscopy (KPFM), which measured local contact potential difference (CPD) between the 2D material and an AFM probe with nanometric spatial resolution [36]. The work function of the 2D materials can be derived from CPD, using the relation $V_{CPD} = \frac{\varphi_{tip} - \varphi_{sample}}{-e}$, where $\phi_{sample}$ and $\phi_{tip}$ are the work functions of the sample and tip, respectively, and e is the electronic charge. This advanced AFM technique offers advantages over commonly used optical measurement techniques, including Raman and photoluminescence spectroscopy [34, 37, 11-12], due to diffraction limited average spot size for collecting local information about the electronic structure of the material. The KPFM line scans of the device (Fig. 3e) shows a clear increase in work function atop the waveguide as compared to the flat unstrained region (corresponding region-B in Fig. 1b). The increment in the work function of MoTe$_2$ ($\Delta$work function=0.08 eV) correspond to change of ~3% tensile strain as calculated by DFT (Fig. 3f). This mismatch in strain variation (with predicted 4% tensile strain) can be attributed to the nature of DFT calculations where MoTe$_2$ is in its isolated form, without considering the effects of substrates and environment [37, 38]. A similar work function change induced by strain and its detection by KPFM is reported for WS$_2$ and graphene [37, 39]. The MoTe$_2$ flakes deform across the waveguide and touch the substrate at different distances away from the waveguide, thus giving rise to asymmetric strain variation. For the detector performance, maximize the optical effective mode overlaps with the strained MoTe$_2$ regions of reduced bandgap.

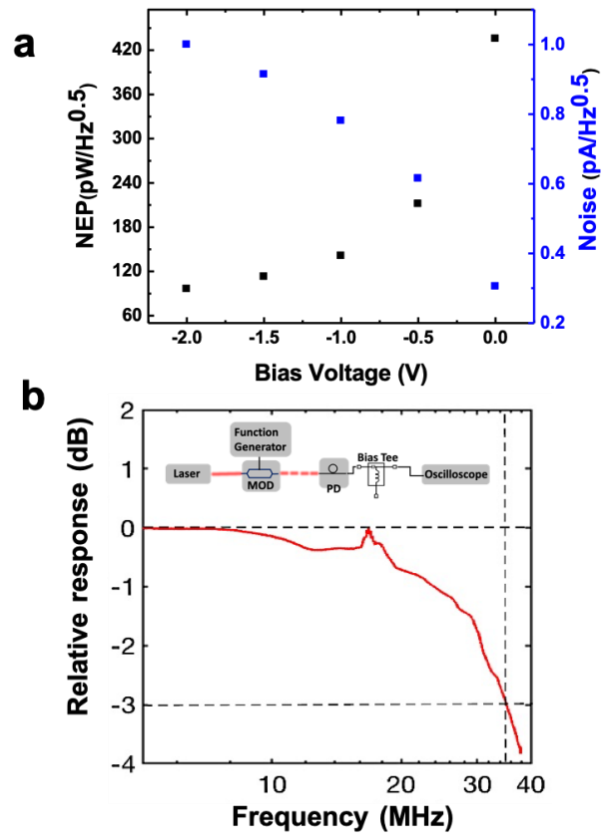

**Figure 4. Dynamic performance** a) The device is operating in photoconductive mode showing low dark current due to strain-lowered bandgap (0.8 eV), which leads to low noise equivalent power of 90 pW/Hz$^{0.5}$ at 2 V, **b)** AC photoresponse as a function of modulated frequency of light signal showing a 3dB cutoff frequency of 35 MHz, which is mainly limited by low transit time. Dynamic response measurement setup (inset).

To understand the detection limit of the device, we determine the noise equivalent power (NEP) i.e., the amount of incident light power that generates a photocurrent equal to the noise current, NEP = $i_n/R$, where, $i_n$ is noise current and $R$ is the responsivity. Generally, at high signal speed, there are mainly two sources of noise; i.e., shot noise ($\sqrt{2qI_d}$) and Johnson noise ($\sqrt{4k_BT/R_{sh}}$), which contribute to the total noise current. However, for a photodetector operating at a photo conducting mode, shot noise always dominates over Johnson noise. Here, NEP is found to be ~ 90 pW/Hz$^{0.5}$, lowest among the devices, although device 2 shows higher responsivity (high dark current) thus revealing the trade-off between sensitivity and noise current. However, our device shows ~2 orders higher sensitivity than graphene/Si photodetector [40] due to lower dark current when operating at photoconductive mode, and comparable sensitivity to NEP to BP photodetectors [41]. The variation of NEP shows a gradual decrease for higher bias voltage (Fig. 4a), enabling low light level sensing, which can be further improved by the formation of a p-n junction.

We perform a dynamic response test of the detector using a modulated laser input (Fig. 4b). The modulated optical output is coupled into the device where the electrical output was measured through a radio frequency microwave (G-S) probe, and the normalized frequency response analyzed via the S$_{21}$ parameter of the network analyzer (Fig. 4b). Our photodetector device displays a 3dB bandwidth of 35.6 MHz at 2 V. The response time of a cavity integrated photodetector is mainly governed by these three factors: i) carrier transit time ($\tau_{tr}$), ii) charge/discharge time of the junction capacitance ($\tau_{RC}$) and iii) photon lifetime ($\tau_{cav}$). Hence, the temporal response of the detector is determined by, $\tau_R = \sqrt{(\tau_{tr}{}^2 + \tau_{RC}{}^2 + \tau_{cav}{}^2)}$. Here, the transit time is given by, $\tau_{tr} = l^2/2\mu V_{bias}$, where $l$ is the channel length of the MoTe$_2$ detector and $\mu$ is the carrier mobility. From top gated FET configuration, when the 2D nanocrystal is atop the waveguide, the field-effect mobility is 1.2 cm$^2$/Vs (Supplementary online information, section 9, Fig. S9). With a channel length of 0.8 μm, the transit time is found to be 3.2 ns.

**Conclusions**

In conclusion, we demonstrate a strain-induced absorption-enhanced 2D nanocrystal (MoTe$_2$) silicon photonic microring-integrated photodetector featuring high responsivity of 0.01 A/W (device 1) and ~0.5 A/W (device 2) at 1550 nm, with a low NEP of 90 pW/Hz$^{0.5}$. Subject to mechanical strain, the bandgap of MoTe$_2$ shifts from 1.04 eV for unstrained to 0.80 eV for strained, when the 2D nanocrystal is wrapped around a non-planarized silicon waveguide. The local enhancement of the work function mapped out by KPFM, confirms a local change of electronic structure of the material due to strain. The device responsivity can be further improved using a high-Q cavity resonator. We observe 3dB bandwidth of 35 MHz, where the response time is transit time-limited. This strain engineered bandgap enables optical absorption at 1550 nm, resulting in an integrated photonic detector that could potentially open up a new pathway for future on-chip photonic circuits.

## Methods

### Density Function Theory (DFT) study

The first-principles density functional theory (DFT) calculations were performed as implemented in the Vienna *ab initio* simulation package (VASP) [42]. The exchange correlation energy is described by the generalized gradient approximation (GGA) using the Perdew-Burke-Ernzerhof (PBE) functionals [43]. Since DFT usually underestimates the bandgaps of semiconductors, we used the HSE06 hybrid functional [44] for the exchange-correlation term, which gives reliable results for bandgaps. The plane-wave cutoff energy was set to 300 eV. Spin-orbit coupling was taken into consideration. The bulk crystal structure was relaxed until the total energy converged to $10^{-6}$ eV and the Hellmann-Feynman force on each atom was less than 0.001 eV/Å. For relaxation, the Brillouin zone was sampled using a Monkhorst-pack 21 x 21 x 21 grid [45]. The optimized lattice parameters for the pristine bulk structure in 2H phase is a= 3.52Å and c=13.97Å. In order to calculate work functions, a large vacuum spacing of 30 Å was added along stacking (out-of-plane) direction. Then we tracked the plane-averaged electrostatic potential into the vacuum, which usually the vacuum energy is reached within a few Å from the surface. Therefore, work function is obtained by subtracting mid-gap energy from the vacuum energy. Here, we used mid-gap energy instead of Fermi energy, since Fermi energy is a quantity that varies with doping and also can be anywhere in the band gap for our cases.

### Device Fabrication

Silicon photonic platform, which includes microring resonator, was fabricated using SOI substrate with a thin 220nm Si layer with 2 µm buried oxide. The devices were patterned using electron-beam lithography using a negative resist ARN 7520. The patterned features were further etched using an Inductively Coupled Plasma (ICP) etching tool. The photoresist mask was removed using acetone wash and short oxygen plasma cleaning. The sample was cleaned using acetone and then Iso-propanol (IPA) and dried using nitrogen gas. The sample was heated on a hot plate for 2-3 mins at 180°C for better adhesion of the flakes. The few-layered MoTe$_2$ flakes were transferred onto the microring resonators(MRRs) using 2D printer tool [30]. After the transfer process a thin layer of PMMA (~300 nm) was spin coated on the sample at 1500 rpm and followed by a post baked process at 180°C for two minutes. For electrical characterization, the metal contact pads (80x 80 $\mu m^2$) were patterned using an EBL process. The sample was then developed in MIBK: IPA (3:1) solution. After development, A mild oxygen plasma was applied to clean the PMMA residues on the exposed 2D layer. A thin layer titanium (5 nm) and gold (45 nm) was deposited using an electron beam evaporation process followed by lift-off in acetone. Details of the step by step fabrication process flow can be found in the supplementary online information (S1, Fig. S1).

**Device Measurement**

To measure the responsivity of the device, we coupled 1,550 nm c.w. input laser and detected the photocurrent through a source meter (Keithly 2600B). Here, $P_{input}$ is the power reaching the MRR-MoTe$_2$ detector, estimated by considering the input grating coupler coupling loss and the silicon waveguide transmission loss (Supplementary online information, Section 10, Fig.S10). The spectral response of the MRR integrated device was measured using a broadband laser (AEDFA-PA-30-B-FA) injected into the grating coupler optimized for the TM mode propagation [46]. The light output from the MRR was coupled to the output fiber and detected by the optical spectral analyzer (OSA202) [47]. The experimental setup for measuring the photodetector devices is shown in Fig. S10 (Supplementary online information).

**KPFM Measurement**

An advanced scanning probe microscopy technique, commonly known as Kelvin Probe Force Microscopy (KPFM) was exploited to gather local electrical information from the sample. For this study, a commercial AFM, MFP-3D (Asylum Research), was used and the probe used was a Budget Sensor Cr/Pt conductive coated tip (Multi75E-G). In order to calculate the work function of the tip, the tip was calibrated by scanning a freshly cleaved highly ordered pyrolytic graphite (HOPG) (φ=4.6 eV), several times and then an average was performed for all the scans. The scanning was done in the attractive (noncontact) imaging regime to avoid any possible contamination of the AFM tip. The work function of the tip was calculated to be 4.1875 eV. In all measurements, the sample was grounded to avoid the possibility of surface charge modifying the CPD value.

**Application Discussion**

Photodetectors are universal building blocks for i) data links and interconnects including hybrid-technology options [48] including dual-function optoelectronic devices [49], ii) photonic integrated network-on-chip technology [50], iii) photonic digital-to-analog converters [51], and photonic analog processors such as iv) photonic residue arithmetic processors relying on cross-bar routers [52, 53], v) analog partial differential equation solvers [54], vi) photonic integrated neural networks [55-58], and vii) photonic reprogrammable circuits such as content-addressable memories [59].

## Acknowledgment

V.S. is supported by AFOSR (FA9550-17-1-0377) and ARO (W911NF-16-2-0194).

## Reference

1.  Jacobsen, R. S. *et al.* Strained silicon as a new electro-optic material. *Nature* **441,** 199–202 (2006).
2.  Cheng, T.-H. *et al.* Strain-enhanced photoluminescence from Ge direct transition. *Appl. Phys. Lett.* **96,** 211108 (2010).
3.  Feng, J., Qian, X., Huang, C.-W. & Li, J. Strain-engineered artificial atom as a broad-spectrum solar energy funnel. *Nat. Photonics* **6,** 866 (2012).
4.  Lee, M. L., Fitzgerald, E. A., Bulsara, M. T., Currie, M. T. & Lochtefeld, A. Strained Si, SiGe, and Ge channels for high-mobility metal-oxide-semiconductor field-effect transistors. *J. Appl. Phys.* **97,** 11101 (2005).
5.  Yun, W. S., Han, S. W., Hong, S. C., Kim, I. G. & Lee, J. D. Thickness and strain effects on electronic structures of transition metal dichalcogenides: 2H-MX 2 semiconductors (M = Mo, W; X = S, Se, Te). *Phys. Rev. B - Condens. Matter Mater. Phys.* **85,** 1–5 (2012).
6.  He, K., Poole, C., Mak, K. F. & Shan, J. Experimental Demonstration of Continuous Electronic Structure Tuning via Strain in Atomically Thin MoS2. *Nano Lett.* **13,** 2931–2936 (2013).
7.  Yue, Q. *et al.* Mechanical and electronic properties of monolayer MoS2 under elastic strain. *Phys. Lett. A* **376,** 1166–1170 (2012).
8.  Ghorbani-Asl, M., Borini, S., Kuc, A. & Heine, T. Strain-dependent modulation of conductivity in single-layer transition-metal dichalcogenides. *Phys. Rev. B - Condens. Matter Mater. Phys.* **87,** 1–6 (2013).
9.  Shi, H., Pan, H., Zhang, Y. W. & Yakobson, B. I. Quasiparticle band structures and optical properties of strained monolayer MoS2 and WS2. *Phys. Rev. B - Condens. Matter Mater. Phys.* **87,** 1–8 (2013).
10. Manzeli, S., Allain, A., Ghadimi, A. & Kis, A. Piezoresistivity and Strain-induced Band Gap Tuning in Atomically Thin MoS2. *Nano Lett.* **15,** 5330–5335 (2015).
11. Castellanos-Gomez, A. *et al.* Local Strain Engineering in Atomically Thin MoS2. *Nano Lett.* **13,** 5361–5366 (2013).
12. Aslan, O. B. *et al.* Probing the Optical Properties and Strain-Tuning of Ultrathin Mo1−xWxTe2. *Nano Lett.* **18,** 2485–2491 (2018).
13. State-of-the-art photodetectors for optoelectronic integration at telecommunication wavelength . *Nanophotonics* **4,** 277 (2015).
14. Feng, B. *et al.* All-Si Photodetectors with a Resonant Cavity for Near-Infrared Polarimetric Detection. *Nanoscale Res. Lett.* **14,** 39 (2019).
15. Walden, R. H. A review of recent progress in InP-based optoelectronic integrated circuit receiver front-ends. in *GaAs IC Symposium IEEE Gallium Arsenide Integrated Circuit Symposium. 18th Annual Technical Digest 1996* 255–257 (1996). doi:10.1109/GAAS.1996.567881

16.     V. K. Narayana, S. Sun, A.-H. Badawya, V. J. Sorger & T. El-Ghazawi, MorphoNoC: Exploring the design space of a configurable hybrid NoC using nanophotonics. *Microprocess. Microsyst.* **50,** 113–126 (2017).

17.     Liu, K., Sun, S., Majumdar, A. & Sorger, V. J. Fundamental Scaling Laws in Nanophotonics. *Sci. Rep.* **6,** 37419 (2016).

18.     Michel, J., Liu, J. & Kimerling, L. C. High-performance Ge-on-Si photodetectors. *Nat. Photonics* **4,** 527 (2010).

19.     Goykhman, I., Desiatov, B., Khurgin, J., Shappir, J. & Levy, U. Waveguide based compact silicon Schottky photodetector with enhanced responsivity in the telecom spectral band. *Opt. Express* **20,** 28594 (2012).

20.     Wang, J. & Lee, S. Ge-photodetectors for Si-based optoelectronic integration. *Sensors* **11,** 696–718 (2011).

21.     Youngblood, N., Chen, C., Koester, S. J. & Li, M. Waveguide-integrated black phosphorus photodetector with high responsivity and low dark current. *Nat. Photonics* **9,** 247 (2015).

22.     Bie, Y. Q. *et al.* A MoTe2-based light-emitting diode and photodetector for silicon photonic integrated circuits. *Nat. Nanotechnol.* **12,** 1124–1129 (2017).

23.     Octon, T. J., Nagareddy, V. K., Russo, S., Craciun, M. F. & Wright, C. D. Fast High-Responsivity Few-Layer MoTe2 Photodetectors. *Adv. Opt. Mater.* **4,** 1750–1754 (2016).

24.     Youngblood, N. & Li, Mo. Integration of 2D materials on a silicon photonics platform for optoelectronics applications . *Nanophotonics* **6,** 1205 (2017).

25.     Ma, P. *et al.* Fast MoTe2 Waveguide Photodetector with High Sensitivity at Telecommunication Wavelengths. *ACS Photonics* **5,** 1846–1852 (2018).

26.     Gan, X. *et al.* Chip-integrated ultrafast graphene photodetector with high responsivity. *Nat. Photonics* **7,** 883 (2013).

27.     Schuler, S. *et al.* Controlled Generation of a p–n Junction in a Waveguide Integrated Graphene Photodetector. *Nano Lett.* **16,** 7107–7112 (2016).

28.     Island, J. O., Steele, G. A., van der Zant, H. S. J. & Castellanos-Gomez, A. Environmental instability of few-layer black phosphorus. *2D Mater.* **2,** 11002 (2015).

29.     Deng, S., Sumant, A. V & Berry, V. Strain engineering in two-dimensional nanomaterials beyond graphene. *Nano Today* **22,** 14–35 (2018).

30.     Johari, P. & Shenoy, V. B. Tuning the Electronic Properties of Semiconducting Transition Metal Dichalcogenides by Applying Mechanical Strains. *ACS Nano* **6,** 5449–5456 (2012).

31.     Shiue, R.-J. *et al.* High-Responsivity Graphene–Boron Nitride Photodetector and Autocorrelator in a Silicon Photonic Integrated Circuit. *Nano Lett.* **15,** 7288–7293 (2015).

32.     Li, H. *et al.* Optoelectronic crystal of artificial atoms in strain-textured molybdenum disulphide. *Nat. Commun.* **6,** 7381 (2015).

33.     R. A. Hemnani, C. Carfano, J. P. Tischler, M. H. Tahersima, R. Maiti, L. Bartels, R. Agarwal, V. J. Sorger, "Towards a 2D Printer: A Deterministic Cross Contamination-free Transfer Method for Atomically Layered Materials", ***2D Materials***: 6, 015006 (2018).

34.     Melitz, W., Shen, J., Kummel, A. C. & Lee, S. Surface Science Reports Kelvin probe force microscopy and its application. *Surf. Sci. Rep.* **66,** 1–27 (2011).

35.     Conley, H. J. *et al.* Bandgap Engineering of Strained Monolayer and Bilayer MoS2. *Nano Lett.* **13,** 3626–3630 (2013).

36.     McCreary, A. *et al.* Effects of Uniaxial and Biaxial Strain on Few-Layered Terrace Structures

of MoS2 Grown by Vapor Transport. *ACS Nano* **10,** 3186–3197 (2016).

37. Sarwat, S. G. *et al.* Revealing Strain-Induced Effects in Ultrathin Heterostructures at the Nanoscale. *Nano Lett.* **18,** 2467–2474 (2018).

38. Jung, D. Y., Yang, S. Y., Park, H., Shin, W. C., Oh, J. G., Cho, B. J., Choi, S.-Y. Interface Engineering for High Performance Graphene Electronic Devices. *Nano Converg.* **2,** 11 (2015).

39. Meng, L. *et al.* Two dimensional WS2 lateral heterojunctions by strain modulation. *Appl. Phys. Lett.* **108,** 263104 (2016).

40. Casalino, M. *et al.* Vertically Illuminated, Resonant Cavity Enhanced, Graphene-Silicon Schottky Photodetectors. *ACS Nano* **11,** 10955–10963 (2017).

41. Huang, L. *et al.* Waveguide-Integrated Black Phosphorus Photodetector for Mid-Infrared Applications. *ACS Nano* **13,** 913–921 (2019).

42. Kresse, G. & Furthmüller, J. Efficient iterative schemes for ab initio total-energy calculations using a plane-wave basis set. *Phys. Rev. B - Condens. Matter Mater. Phys.* **54,** 11169–11186 (1996).

43. Perdew, J. P., Burke, K. & Ernzerhof, M. Generalized gradient approximation made simple. *Phys. Rev. Lett.* **77,** 3865–3868 (1996).

44. Paier, J. *et al.* Screened hybrid density functionals applied to solids. *J. Chem. Phys.* **124,** 154709 (2006).

45. Pack, J. D. & Monkhorst, H. J. 'special points for Brillouin-zone integrations'-a reply. *Phys. Rev. B* **16,** 1748–1749 (1977).

46. R. Maiti, C. Patil, R. Hemnani, M. Miscuglio, R. Amin, Z. Ma, R. Chaudhary, A. T. C. Johnson, L. Bartels, R. Agarwal, V. J. Sorger, "Loss and Coupling Tuning via Heterogeneous Integration of MoS$_2$ Layers in Silicon Photonics" *Optics Materials Express*, **9,** 2, 751-759 (2018).

47. R. Maiti, R. A. Hemnani, R. Amin, Z. Ma, M.Tahersima, T. A. Empante, H. Dalir, R. Agarwal, L. Bartels, V. J. Sorger, "A semi-empirical integrated microring cavity approach for 2D material optical index identification at 1.55 um" *Nanophotonics* **8(3),** 435-441 (2019)**.**

48. S. Sun, A. Badaway, V. Narayana, T. El-Ghazawi, V. J. Sorger "Photonic-Plasmonic Hybrid Interconnects: Efficient Links with Low latency, Energy and Footprint" *IEEE Photonics Journal* 7, 6 (2015).

49. S. Sun, R. Zhang, J. Peng, V. K. Narayana, H. Dalir, T. El-Ghazawi, V. J. Sorger "MODetector (MOD): A Dual-Function Transceiver for Optical Communication On-Chip" *Optics Express* 57, 18, 130-140 (2018).

50. V. K. Narayana, S. Sun, A.-H. Badawya, V. J. Sorger, T. El-Ghazawi "MorphoNoC: Exploring the Design Space of a Configurable Hybrid NoC using Nanophotonics*" Microprocessors and Microsystems* 50, 113-126. (2017).

51. J. Meng, M. Miscuglio, J. K. George, V. J. Sorger "Electronic Bottleneck Suppression in Next-generation Networks with Integrated Photonic Digital-to-analog Converters*"arXiv* preprint: 1911:02511 (2019).

52. S. Sun, V. K. Narayana, I. Sarpkaya, J. Crandall, R. A. Soref, H. Dalir, T. El-Ghazawi, V. J. Sorger "Hybrid Photonic-Plasmonic Non-blocking Broadband 5×5 Router for Optical Networks*"IEEE Photonics Journal* 10, 2 (2018).

53. J. Peng, S. Sun, V. Narayana, V.J. Sorger, T. El-Ghazawi "Integrated Nanophotonics

Arithmetic for Residue Number System Arithmetic" *Optics Letters* 43, 9, 2026-2029 (2018).

54. S. Sun, M. Miscuglio, R. Zhang, Z.Ma, E. Kayraklioglu, C. Chen, J. Crandall, J. Anderson, Y. Alkabani, T. El-Ghazawi, V. J. Sorger "Analog Photonic Computing Engine as Approximate Partial Differential Equation Solver" *arXiv* preprint: 1911.00975 (2019).

55. M. Miscuglio, A. Mehrabian, Z. Hu, S.I. Azzam, J.K. George, A.V. Kildishev, M. Pelton, V.J. Sorger "All-optical Nonlinear Activation Function for Photonic Neural Networks" *Optical Material Express* 8(12), 3851-3863 (2018).

56. J. K. George, A. Mehrabian, R. Armin, J. Meng, T. Ferreira De Lima, A. N. Tait, B. Shastri, P. Prucnal, T. El-Ghazawi, V. J. Sorger "Noise and Nonlinearity of Electro-optic Activation Functions in Neuromorphic Compute Systems" *Optics Express* 27, 4 (2019).

57. A. Mehrabian, M. Miscuglio, Y. Alkabani, V. J. Sorger, T. El- Ghazawi "A Winograd-based Integrated Photonics Accelerator for Convolutional Neural Networks" *IEEE J. of Selected Topics in Quantum Electronics* 26(1), 1-12 (2019).

58. M. Miscuglio, G.C. Adam, D. Kuzum, V.J. Sorger "Roadmap on Material-Function Mapping for Photonic-Electronic Hybrid Neural Networks" *APL Materials* 7, 100903 (2019)

59. Y. Alkabani, M. Miscuglio, V.J. Sorger, T. El-Ghazawi "OE-CAM: A Hybrid Opto-Electronic Content Addressable Memory" *IEEE Photonics Journal arXiv* preprint: 1912:02221 (2019).